\documentclass[twocolumn]{jpsj2} 
\usepackage{epsfig,psfrag,amsmath,amssymb,float,color}
\title{Confinement and critical regime in doped frustrated quasi-one dimensional magnets}

\author{
\textsc{Nicolas Laflorencie}$^{1,2}$, \textsc{Didier Poilblanc}$^{1}$
}

\inst{$^{1}$ Laboratoire de Physique Th\'eorique, CNRS-UMR5152
Universit\'e Paul Sabatier, F-31062 Toulouse, France \\
$^{2}$ Department of Physics \& Astronomy, University of British Columbia,  
Vancouver, B.C., Canada, V6T 1Z1\\}

\abst{Ground state and finite temperature properties of a system of coupled
frustrated and/or dimerized spin-1/2 chains modeling e.g. the CuGeO$_3$ compound are reviewed.
Special emphasis is put on the investigation of the role of impurity doping. A chain-mean field computation
combining exact diagonalisations of the chain hamiltonians together with a 
mean field treatment of the weak interchain couplings is performed in order 
to map the microscopic model onto a low-energy effective model. The latter describes
a 2-dimensional system of effective spin-1/2 local moments interacting by spacially anisotropic 
long range spin exchange interactions. 
An extensive study of this effective model is performed by
Stocastic Series Expansion Quantum Monte Carlo for a wide range of temperatures and impurity concentrations.
Interesting scaling behaviors of the uniform and staggered spin susceptibilities (above a small 
ordering N\'eel temperature due to a residual 3D coupling) can be interpreted
in terms of the formation of large clusters of correlated spins carrying a finite magnetization.  
Such results are reproduced satisfactorily by a new Real Space RG 
enabling to deal with long range interactions in 
two-dimensions.}

\kword{Strongly correlated systems, Quantum magnetism, Critical behaviors, Numerical simulations, Random magnets}

\begin{document}
\newcommand{\bc}{\begin{center}}
\newcommand{\ec}{\end{center}}
\newcommand{\be}{\begin{equation}}
\newcommand{\ee}{\end{equation}}
\newcommand{\beqn}{\begin{eqnarray}}
\newcommand{\eeqn}{\end{eqnarray}}
\def\1.2{\frac{1}{2}}
\def\AF{antiferromagnetic~}
\def\AFp{antiferromagnetic.~}
\def\AFv{antiferromagnetic,~}
\maketitle

\section{Introduction}
Low dimensional gapped quantum magnets have attracted a lot of interest in condensed matter physics for many years. The possibility of doping such systems has lead to an extremely rich emerging field. For instance, the discovery of the first non-organic spin-Peierls (SP) compound CuGeO$_3$ \cite{Hase93} and its doping with static non-magnetic impurities realized by direct substitution of a small fraction of copper atoms by zinc \cite{Zn_CuGeO3} or magnesium \cite{Mg_CuGeO3} atoms offered a new challenge for the theorist. Replacing a spin-$\1.2$ in a {\it spontaneously} dimerized spin chain by a non-magnetic impurity, described as an inert site, releases a free spin-$\1.2$, named a soliton, which does not bind to the dopant \cite{Sorensen98}. The physical picture is
completely different when a {\it static} bond dimerisation exists
and produces an attractive potential between the soliton and the
dopant~\cite{Sorensen98,Nakamura99} and consequently leads, under
doping, to the formation of local magnetic
moments~\cite{Sorensen98,Normand2002} as well as a rapid
suppression of the spin gap~\cite{Martins96}. However, a coupling
to a purely one-dimensional (1D) adiabatic lattice~\cite{Hansen99}
does not produce confinement in contrast to more realistic models
including an elastic inter-chain coupling (to mimic 2D or 3D
lattices)~\cite{Hansen99,Dobry99}. 

In the following, we are going to present through a generic model 
of quasi one-dimensional SP system the impurity induced local moment formation as well as the tendency 
towards antiferromagnetic (AF) ordering in the thermodynamic limit observed for any non-zero dopant concentration. An interesting low temperature 
scaling behavior
will also be presented and interpreted in term of the formation of large clusters of correlated spins carrying a finite magnetization. This short review is the consequence of several previous works~\cite{ourprl,Effective04,LPS05} involving many numerical tools: Lanczos Exact Diagonalization (ED) associated to chain-mean field theory, Stocastic Series Expansion (SSE) Quantum Monte Carlo (QMC) methods including long-range interactions, numerical Real Space Renormalization Group (RSRG) technique.

The rest of the paper is organized as follows. In the following section, we introduce a generic microscopic model describing weakly coupled frustrated 
spin-$\frac{1}{2}$ chains, and compute its phase diagram without impurity. Then in section 3 we concentrate on impurity efects, through the 
confinement mechanism, responsible of the impurity induced local moment formation. We also study the effective interaction between impurities which 
enable us to construct an effective two-dimensional (2D) diluted model. The section 4 is thus devoted to the impurity induced AF 
ordering which appear in such an effective model. 
Finally, in the section 5 we focus on the low temperature regime of the effective 2D model for which a RSRG procedure has been developed.
\section{Microscopic model of coupled frustrated spin-$\1.2$ chains}
\subsection{Generic Hamiltonian for weakly coupled spin-Peierls chains}
We start with the microscopic Hamiltonian 
\begin{eqnarray} 
\label{hamilQ1D.MF}  
{\cal{H}}=\sum_{a=1}^{M}\sum_{i=1}^{L}[J(1+\delta_{i,a}){\vec S}_{i,a}\cdot{\vec S}_{i+1,a}+\alpha J{\vec S}_{i,a}\cdot {\vec S}_{i+2,a} \nonumber\\
+h_{i,a}S_{i,a}^z], 
\end{eqnarray} 
as a generic model for a quasi one-dimensional frustrated spin system \cite{ourprl} with $J$ and $\alpha$ both positive, and $S=\1.2$. $i$ is a lattice index along the chain of size $L$ and $a$ labels the $M$ chains. 
\begin{figure}[!ht] 
\bc
\psfrag{Y}{{{$J_1$}}}  
\psfrag{X}{{\color{red}{$\alpha J_1$}}} 
\psfrag{Z}{{\color{green}{$J_{\bot}$}}}  
\psfrag{U}{{\color{blue}{$J_{4}$}}}  
\epsfig{file=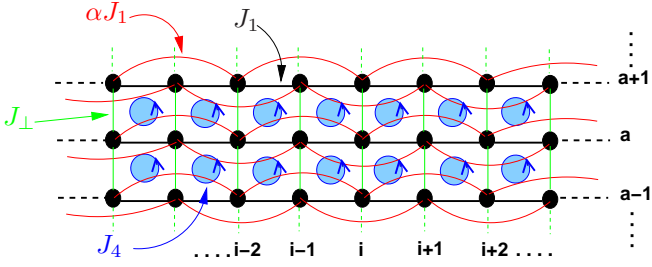,width=\columnwidth,clip} 
\caption{Schematic picture of a system of
coupled frustrated spin-$\1.2$ chains governed by Eq.~(\ref{hamilQ1D.MF}).} 
\label{fig:LatticeJ4} 
\ec 
\end{figure} 
The effective magnetic field $h_{i,a}$ results from a mean-field (MF) treatment \cite{Schulz96} of a weak interchain coupling, $J_{\perp}{\vec S}_{i,a}\cdot {\vec S}_{i,a+1}$ with $J_{\perp}\ll J$, computed in a self-consistent way
\be \label{selfH} h_{i,a}=J_{\bot}(\langle S_{i,a+1}^z \rangle + \langle S_{i,a-1}^z \rangle). \ee 
The modulation of the nearest neighbor exchange $\delta_{i,a}$ is also computed self-consistently using
\be \label{self}  \delta_{i,a}=\frac{J_4}{J_1}(\langle {\vec S}_{i,a+1}\cdot{\vec S}_{i+1,a+1} \rangle + \langle {\vec S}_{i,a-1}\cdot{\vec S}_{i+1,a-1} \rangle). \ee 
This modulation term might have in fact multiple origins: although a four-spin cyclic exchange mechanism provides the most straightforward derivation of it \cite{ourprl}, at a qualitative level,   $J_4$ can also mimic higher-order effects in $J_{\perp}$ \cite{Byrnes99} or the coupling to adiabatic phonons \cite{Dobry99}. Indeed, in that case, due to a magnetoelastic coupling, the modulations $\delta_{i,a}$ result from small displacements of the ions. The elastic energy is the sum of a local term $\frac{1}{2}K_\parallel\sum_{i,a}\delta_{i,a}^2$ and an interchain contribution $K_\perp\sum_{i,a}\delta_{i,a}\delta_{i,a+1}$ of electrostatic origin~\cite{AffleckElast}. Then, the mean-field equation (\ref{self}) is replaced by \cite{Hansen99} %
\begin{equation} \label{self3} K_\parallel\delta_{i,a}+K_\perp(\delta_{i,a+1}+\delta_{i,a-1})= J_1\langle {\vec S}_{i,a}\cdot{\vec S}_{i+1,a} \rangle , \end{equation} %
giving very similar results \cite{note1} so that we can restrict ourselves to Eq.~(\ref{self}) without loss of generality.
%
\begin{figure}[!ht] 
\begin{center} \psfrag{A1}{{\small{$J_4=0.01$}}} \psfrag{A2}{{\small{$J_4=0.05$}}} \psfrag{A3}{{\small{$J_4=0.1$}}} \psfrag{A4}{{\small{$J_4=0.2$}}} \psfrag{A5}{$J_{\bot}$} \psfrag{AF}{{\Large{\bf{AF}}}} \psfrag{SP}{{\Large{\bf{SP}}}} \psfrag{ac}{\small{$\alpha_c$}} \psfrag{alp}{$\alpha$} \psfrag{L12}{\tiny{$L=12$}} \psfrag{L16}{\tiny{$L=16$}} \psfrag{F S E}{{\small {F S E}}} \epsfig{file=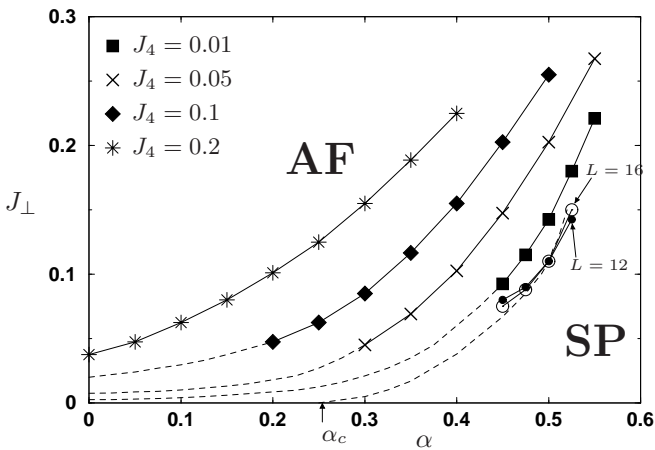,width=\columnwidth,clip}  \caption{SP-AF phase diagram in
the ($\alpha,J_{\bot}$) plane from ED of chains of length up to
$16$ sites. Symbols correspond to different values of $J_4\ge 0$
as indicated on plot. Typically, finite size effects (FSE) are smaller than the size of
the symbols. The computed transition lines are extended by {\it
tentative} transition lines (dashed lines) in the region where FSE
become large. At $J_4=0$ we have plotted a few points in the
vicinity of the MG point for $L=12$ and $L=16$. (Figure reprinted from Ref.\cite{ourprl}).}  
\label{fig:PhDg} 
\end{center} 
\end{figure} 
%
We first consider the clean undoped system where the $M$ chains becomes equivalent (see Fig.~\ref{fig:LatticeJ4}). By solving the self-consistency conditions Eqs.~(\ref{selfH}-\ref{self}) using ED on finite chains with up to 16 sites~\cite{NoteLNP} (supplemented by a finite size scaling analysis), the transition lines $J_{\perp}=J_{\perp}(\alpha,J_4)$ separating the dimerized SP phase ($h_{i,a}=0$) and the AF ordered phase ($h_{i,a}\ne 0$) have been obtained \cite{ourprl,LectureNotes04}.
The phase diagram is shown in Fig. \ref{fig:PhDg}.
If the chains are not dimerized, $\delta
J_{i,a}$ is constant and the exchange is just renormalised. On the
other hand, if the SP phase is stable, each chain displays the
same dimerized pattern when $J_4<0$ whereas dimer order is
staggered in the transverse direction when $J_4>0$. Note that,
apart from special features (see below), physical properties for
positive or negative $J_4$ are quite similar. Note also that the
dimerized GS would be $2^M$-fold degenerate when $J_4=0$ (each
chain is independently two-fold degenerate) while the degeneracy
is reduced to 2 when $J_4 \neq 0$. When $\alpha=J_{\bot}=0$ but
$J_4\neq 0$ each chain spontaneously dimerizes and a gap opens up.
Consequently, a finite value of the AF inter-chain coupling is
necessary to drive the system into the AF ordered
phase~\cite{Matsumoto02}. The frustration $\alpha$ stabilizes
further the dimerized phase with respect to the AF one, the
critical $J_\perp(\alpha)$ increasing with increasing $\alpha$ as
seen from the phase diagram shown in Fig.~\ref{fig:PhDg}.
\section{Impurity effects}
Let us now turn to the doped case. A non-magnetic dopant is
described here as an inert site decoupled from its neighbors.
Under doping the system becomes
non-homogeneous so that the MF equations
are solved self-consistently on finite $L\times M$ clusters and
lead to a non-uniform solution. At each step of the MF iteration
procedure, we use Lanczos ED techniques to treat {\it exactly}
(although independently) the $M$ {\it non-equivalent} finite
chains  and compute
$\langle S^{z}_{i,a}\rangle$ for the next iteration step
until the convergence is eventually achieved.
\subsection{One impurity:~soliton confinement}
\begin{figure}[!ht]
\begin{center}
\psfrag{12}{$L=12$} \psfrag{L16}{$L=16$}
\psfrag{S}{{\small{$\langle S_{i,1}^z \rangle$}}}
\psfrag{z}{{\small{$i$}}} \psfrag{J4}{$J_{4}$} \psfrag{xi}{$\xi_{\parallel}$}
\psfrag{lo}{{\small{$2\xi_{\parallel}$}}} 
\epsfig{file=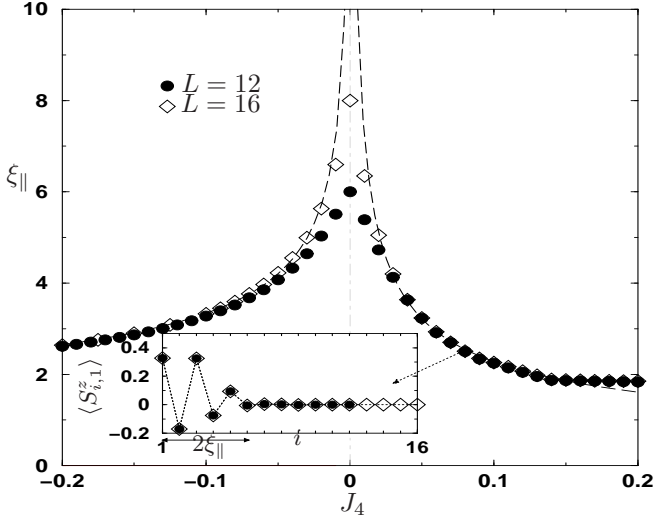,width=\columnwidth,height=7cm,clip}
\caption{ED data of the soliton average position  vs $J_4$
calculated for $\alpha=0.5$ and $J_{\bot}=0.1$. Different symbols
are used for $L\times M$ = $12\times 6$ and $16\times 8$ clusters.
The long-dashed line is a power-law fit (see text). Inset shows
the magnetization profile in the doped ($a=1$) chain at $J_4=0.08$, ie
$\xi_{\parallel} \simeq 2.5$ (Figure reprinted from Ref.\cite{ourprl}).}
\label{fig:xi}
\end{center}
\end{figure}

We first consider the case of a single dopant. Whereas in the case $J_4 =0$  the soliton remains de-confined, a very small $J_4 \neq 0$ is sufficient to produce a confining string which binds the soliton to the dopant \cite{ourprl}. 
In order to describe quantitatively the confinement mechanism near the dopant, we can define a confinement length in the chain direction $\xi_{\parallel}$ as
\begin{equation}
\xi_{\parallel}=\frac{\sum_{i}i|\langle S_{i}^{z}\rangle|}{\sum_{i}|\langle S_{i}^{z}\rangle|},
\end{equation}
We have calculated it for a $16 \times 8$ system with $\alpha=0.5$ and 
$J_{\perp}=0.1$, and we show its variation as a function of $J_4$ in 
Fig.\ref{fig:xi}. Note that $\xi_{\parallel}(J_4)\neq\xi_{\parallel}(-J_4)$ and a
power law~\cite{Nakamura99} with different exponents $\eta$ is
expected when $J_4\rightarrow 0$. A fit gives $\eta\sim 0.33$ if
$J_4<0$ and $\eta\sim 0.50$ for $ J_4>0$ (Fig.\ref{fig:xi}). This
asymmetry can be understood from opposite renormalisations of
$J_1$ for different signs of $J_4$. Indeed, if $J_4<0$ then
$\delta J_{i,a}>0$ and the nearest neighbor MF exchange becomes
larger than the bare one. Opposite effects are induced by $J_4>0$.
%
\subsection{Two impurities:~effective interaction}
\begin{figure}[!ht]
\bc
\psfrag{A}{{{$\Delta i$}}} 
\psfrag{B}{{{$\Delta a$}}}

\epsfig{file=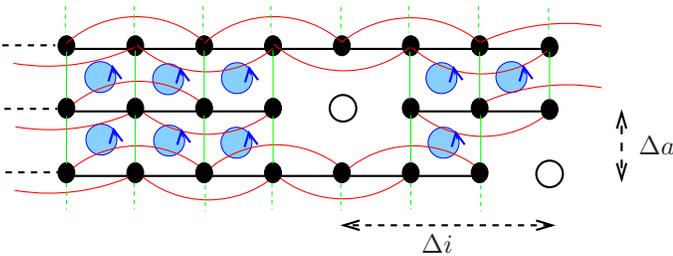,width=\columnwidth,clip}
\caption{Schematic picture of the system of coupled frustrated spin-$\frac{1}{2}$ chains doped with two non magnetic impurities separated by a distance $\Delta i$ (resp. $\Delta a$) in the chains direction (resp. transverse direction).}
\label{fig:LatticeJ4.2Imp}
\ec
\end{figure}

\begin{figure}[!ht]
\bc 
\psfrag{a=0}{$a=0$~~~~(a)}
\psfrag{a=1}{$a=1$~~~~(b)}
\psfrag{a=2}{$a=2$~~~~(c)}
\psfrag{a=3}{$a=3$~~~~(d)}
\epsfig{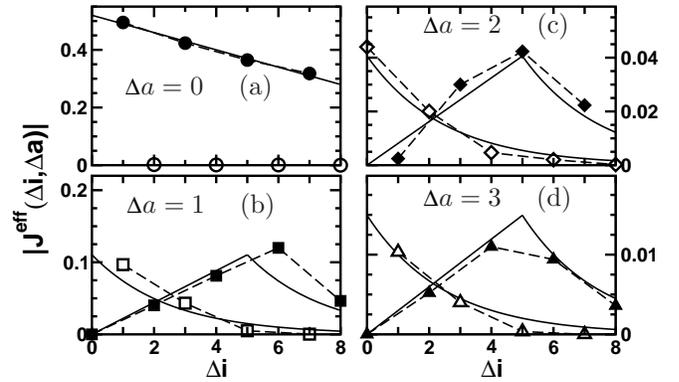} 
\caption{Magnitude of the
effective magnetic coupling between two impurities located either
on the same chain (a)  or on different ones (b-c-d)  vs the dopant separation $\Delta i$ in
a system of size $L \times M=16 \times 8$ with $\alpha=0.5$,
$J_{\bot}=0.1$, and $J_4=0.08$. Closed (resp. open) symbols
correspond to AF (F) interactions. Full lines are fits (see text). (Figure reprinted from Ref.\cite{LectureNotes04}).}
\label{fig:Jeff}
\ec
\end{figure}

We now turn to the investigation of the effective interaction between dopants. 
Each impurity releases an effective spin $\frac{1}{2}$, localized at a distance $\sim \xi_{\parallel}$ from it due to the confining
 potentiel set by $J_4$.
 When two impurities are introduced in the system of coupled chains (see Fig.~\ref{fig:LatticeJ4.2Imp}), we can define an effective pairwise interaction $J^{\rm eff}$ as the
energy difference of the $S=1$ and the $S=0$ ground state (GS). When $J^{\rm
eff}=E(S=1)-E(S=0)$ is positive (negative) the spin interaction is
AF (ferromagnetic). 
%
%
Let us first consider the case of two
dopants in the same chain. (i) When the two vacancies are on the
same sub-lattice the moments experience a very small ferromagnetic
$J^{\rm eff}<0$ as seen in Fig.~\ref{fig:Jeff} with $\Delta a=0$ so that the
two effective spins $\frac{1}{2}$ are almost free. (ii) When the
two vacancies sit on different sub-lattices, $\Delta i$ is odd and
the effective coupling is AF with a magnitude close to the
singlet-triplet gap. Fig.~\ref{fig:Jeff} with $\Delta a=0$ shows that the decay
of $J^{\rm eff}$ with distance is in fact very slow for such a
configuration. The behavior of the pairwise interaction of two dopants located on
{\it different} chains ($\Delta a=1,2,3$) is shown in
Fig.~\ref{fig:Jeff} for  $\Delta a=1,2,3$~for $J_4>0$. When dopants are on
opposite sub-lattices the effective interaction is
antiferromagnetic. At small dopant separation $J^{\rm eff}(\Delta
i)$ increases with the dopant separation as the overlap between
the two AF clouds increases until $\Delta i \sim 2\xi_{\parallel}$. For larger
separation, $J^{\rm eff}(\Delta i)$ decays rapidly. If dopants are on the same sub-lattice,
solitons are located on the same side of the dopants and the effective exchange $J^{\rm eff}(\Delta i)$ is
ferromagnetic and decays rapidly to become negligible when $\Delta i >
2\xi$. The key feature here is the fact that the
effective pairwise interaction is {\it not} frustrating (because
of its sign alternation with distance) although the frustration is
present in the microscopic underlying model.
Our next step is to fit the numerical data in order to derive an analytic expression for $J_{\rm{eff}}$. 
Using only five parameters, two energy scales and three length scales, we can fit ED data with very simple mathematical 
expressions. When $\Delta a=0$ (same chain), $J^{\rm {eff}}$ approximately fulfills $J^{\rm eff}(\Delta i,0)=J_0(1-\Delta i/\xi_{\parallel}^{0})$ 
for $\Delta i$ even and $\Delta i < \xi_{\parallel}^{0}$, and $J^{\rm eff}(\Delta i,0)=0$ otherwise. For dopants located on 
different chains and on the same sub-lattice ($\Delta i+\Delta a$ even) one has,
\be
J^{\rm eff}(\Delta i,\Delta a)=-J^{'}_{0}\exp(-\frac{\Delta i}{\xi_{\parallel}})\exp(-\frac{\Delta a}{\xi_{\perp}}),
\ee
while if the dopants are on opposite sub-lattices, one gets
\be
J^{\rm eff}(\Delta i,\Delta a)=J^{'}_{0}\frac{\Delta i}{2\xi_{\parallel}}\exp(-\frac{\Delta a}{\xi_{\perp}})
\ee
for $\Delta i\le 2\xi_{\parallel}$ and
\be
J^{\rm eff}(\Delta i,\Delta a)=-J^{'}_{0}\exp(-\frac{\Delta i-2\xi_{\parallel}}{\xi_{\parallel}})\exp(-\frac{\Delta a}{\xi_{\perp}}),
\ee
for $\Delta i > 2\xi_{\parallel}$. The fitting parameters are $J_{0}=0\
.52$, $J_{0}^{'}=0.3$, $\xi_{\parallel}^{0}=17.33$, $\xi_{\parallel}=2\
.5$ and $\xi_{\perp}=1$ in the case considered here : $\alpha=0.5, J_{\
\perp}=0.1~{\rm{and}}~J_4 =0.08$ (see Fig.\ref{fig:Jeff}).
\section{Impurity induced antiferromagnetic ordering}
In order to study the impurity induced AF ordering in the system of weakly coupled chains [Eq.~(\ref{hamilQ1D.MF})], a long-range {\it non-frustrated} effective
model of diluted effective spin-$\frac{1}{2}$ is defined,
\be
{\mathcal{H^{\rm{eff}}}}
=\sum_{{\bf r}_1,{\bf r}_2}
\epsilon_{{\bf r}_1} \epsilon_{{\bf r}_2}J^{\rm eff}({\bf r}_1-{\bf r}_2)
{\bf S}_{{\bf r}_1}\cdot {\bf S}_{{\bf r}_2} ,
\label{EffHam}
\ee
with $\epsilon_{\bf r}=1$ ($0$) with probability $x$ ($1-x$),
where $x$ is the dopant concentration. Such a model can be
studied by Quantum Monte Carlo (QMC) simulations on $L_x \times L_y$ clusters much larger than
those accessible to ED~\cite{Dobry99} and {\it at all temperatures}.
\begin{figure}[!ht]
\bc 
\epsfig{file=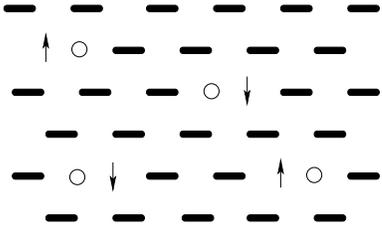,width=5cm,clip} 
\caption{Schematic picture
of a doped SP system. Thick bonds correspond to dimers, and
non-magnetic dopants (released spin-$\frac{1}{2}$)
are represented by open circles (arrows).}
\label{fig:PictEff}
\ec
\end{figure}
%

We study the effective Heisenberg model using the powerfull Stochastic Series Expansion (SSE) method
\cite{Sandvik98} to investigate GS as well as
finite $T$ properties. In this approach, the interactions are sampled
stochastically, and for a long-ranged interaction the computational
effort is then reduced from $\sim N_s^2$ to $N_s\ln{(N_s)}$
\cite{Sandvik03}. In order to accelerate the convergence of the
simulations at the very low temperatures needed to study the ground
state, we use a $\beta$-doubling scheme \cite{SandPerc} where the
inverse temperature is successively increased by a factor $2$.
Comparing results at several $\beta =2^n$, one can subsequently
check that the $T\to 0$ limit has been reached.
\subsection{$T=0$ antiferromagnetic ordering}
\begin{figure}
\bc \epsfig{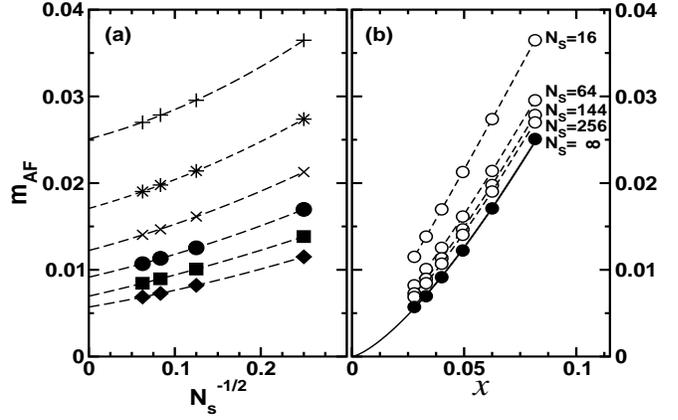} \caption{Staggered
magnetization per site. Disorder average has been done
over at least $2000$ samples. (a) Finite size extrapolations
(see text) at fixed doping $x$. The different symbols are used for different concentrations $x$ with, from top to bottom, 
$x=4/49,~0.0625,~4/81,~0.04,~4/121,~1/36$. (b) Doping dependance
of $m_{\rm AF}$ for various numbers of spins and in the thermodynamic
limit (full symbols). (Figure reprinted from Ref.\cite{Effective04})} \label{fig:mAF} \ec
\end{figure}

The AF ordering instability is signalled by the divergence with
system size of the staggered structure factor,

\be \label{StgStc}
S(\pi,\pi)=\frac{1}{L_xL_y} \langle (\sum_{i} (-1)^{i}
S_{i}^{z})^2\rangle \, .
\ee
Note that within our effective model approach, only the sites carrying a
``dopant spin'' contribute to this sum. It is convenient to normalized 
$S$
with respect to the number of sites, i.e. to define a staggered structure
factor per site; $s(\pi,\pi)=S(\pi,\pi)/L_x L_y$. In an ordered AF state,
$s(\pi,\pi)$ should converge, with increasing size, to a non-zero
value $< 1/4$. The (finite size) sublattice magnetization $m_{\rm AF}$ can
then be obtained by averaging $s(\pi,\pi)$ over a
large number of dopant distributions, i.e.
$(m_{AF})^{2} = 3\langle s(\pi,\pi)\rangle_{\rm {dis}}$, where
the factor 3 comes from the spin-rotational invariance~\cite{Reger88}
and $\langle\ldots\rangle_{\rm {dis}}$ stands for the disorder average.
The staggered magnetization per dopant is then simply
$m_{\rm {spin}}=m_{\rm AF}/x$. 
Since, strictly speaking, in 2D the divergence of $S(\pi,\pi)$
occurs only at $T=0$ ($\beta=\infty$) it is appropriate to first
extrapolate the finite size numerical data to $T=0$, and then using a
polynomial fit in $1/\sqrt{N_{s}}$ an
accurate extrapolation to the thermodynamic limit is performed as shown
in Fig.\ref{fig:mAF}(a). The doping dependance of the extrapolated
$m_{AF}$ is given in Fig.\ref{fig:mAF}(b). We have tested various fits to the data.
Assuming a power law $\propto x^{\mu}$, the best fit (solid line in Fig.\ref{fig:mAF}(b)) gives an
exponent $\mu\simeq 1.38 >1$. 
%
%
\subsection{N\'eel temperature}
\begin{figure}
\bc \epsfig{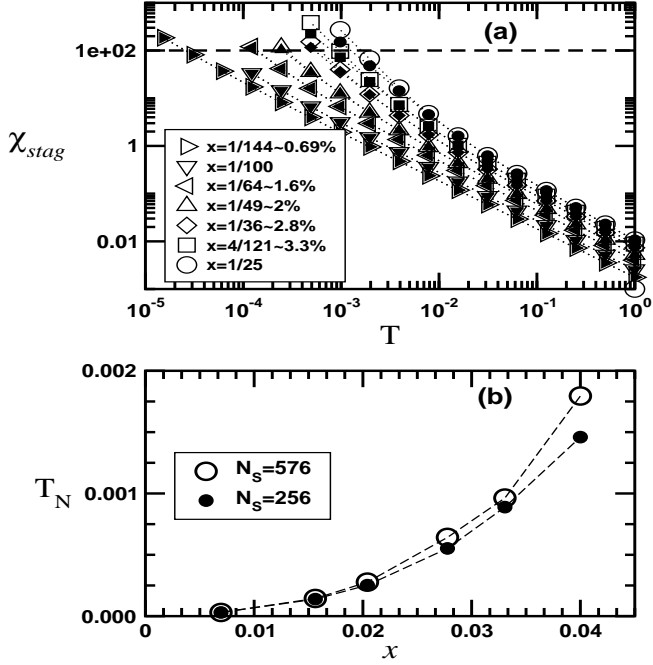} 
\caption{(a)
Staggered susceptibility of a 2D layer vs temperature (using
log-log scales) for $N_s=256$ (full symbols) and $N_s=576$ (open
symbols) spins. Concentrations $x$ are shown on the
plot. $\lambda_{3D}^{-1}=100$ is shown by the dashed line.
(b)
N\'eel temperature vs dopant concentration $x$ for a 3D RPA
inter-plane coupling $\lambda_{3D}=0.01$ and for $N_s=256$ and
$N_s=576$ spins. (Figure reprinted from Ref.\cite{Effective04})} \label{fig:stag} \ec
\end{figure}

It is also very instructive to calculate the N\'eel temp\'erature, asssuming a small (effective) 3D magnetic coupling
$\lambda_{3D}$ between the 2D planes. Using an RPA criterion, the
critical temperature $T_N$ is simply given by $\chi_{\rm
stag}(T_N)=1/|\lambda_{3D}|$ where the staggered spin
susceptibility (normalized per site) is defined as usual by,
\begin{equation}
\label{eq:RPA} \chi_{_{\rm stag}}(T)=\frac{1}{L^2} \sum_{i,j}
(-1)^{r_i + r_j} \int_0^\beta d\, \tau
\langle S_{i}^{z}(0)S_{j}^{z}(\tau)\rangle \, ,
\end{equation}
and averaged over several disorder configurations (typically
$2000$). Since $\chi_{\rm stag}(T_N)$ is expected to reach its
thermodynamic limit for a {\it finite} linear size $L$ as long as
$T_N$ remains {\it finite}, accurate values of $T_N$ can be
obtained using a finite size computation of $\chi_{\rm stag}(T)$
for not too small inter-chain couplings. Fig.~\ref{fig:stag}(a)
shows that $\chi_{\rm stag}(T)$ diverges when $T\to 0$. $T_N$ is
determined by the intersection of the curve $\chi_{\rm stag}(T)$
with an horizontal line at coordinate $1/\lambda_{3D}$. Note that
finite size corrections remain small, even in the worst case
corresponding to very small $\lambda_{3D}$ values and large dopant
concentrations. The doping dependance of $T_N$ is plotted in
Fig.~\ref{fig:stag}(b) for a particularly small value
$\lambda_{3D}=0.01$ (in order to show the small size dependance
observable in that case). It clearly reveals a rapid decrease of
$T_N$ when $x\rightarrow 0$, but, in agreement with experiments,
does not suggest a non-zero critical concentration. In Fig.\ref{fig:stag}(b),
we show the behavior of $T_N(x)$ down to $x\simeq0.007$. Note that from
numerical fits of the data, we can not
clearly distinguish between a power-law behavior (with an exponent $\sim 2.5$)
and an exponential law like $A\exp(-B/x)$, as suggested by fits of experimental
data for Cu$_{1-x}$Zn$_x$GeO$_3$ \cite{Manabe98}.

\section{Low temperature critical properties}
\subsection{Saturation of the Curie constant}
%
\begin{figure}[!ht]
\bc 
\psfrag{Cu}{${\overline{c(T)}}$}
\psfrag{C(T)}{${\overline{c(T)}}$}
\psfrag{T}{$T$}
\psfrag{a}{{{$\frac{1}{12}$}}}
\psfrag{Te}{$T$}
\epsfig{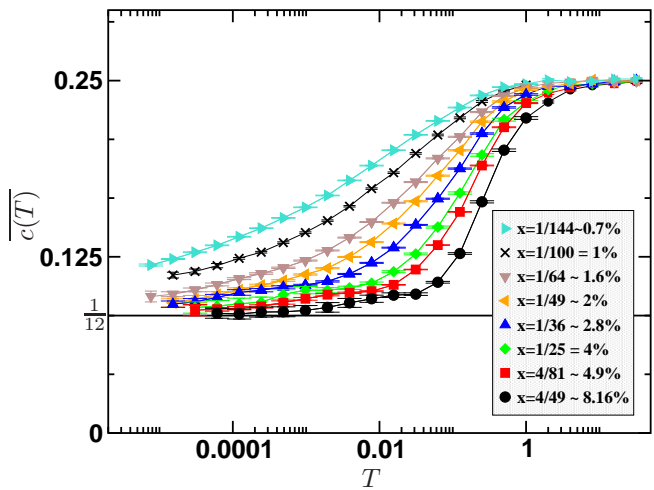} 
\caption{Curie constant per spin $\langle C \rangle$ plotted vs the energy scale. Quantum Monte Carlo SSE
results shown vs $T$ for $N_s =256$ spins and 9 different
concentrations $x$ indicated on the plot. Disorder is performed over
$10^3$ to $10^4$ samples. (Figure reprinted from Ref.\cite{LPS05}).} 
\label{fig:CSSE}
\ec
\end{figure}
%
As a first investigation, we have computed the uniform
susceptibility $\chi(T)$ for a wide range of temperatures using SSE simulations.
Results for the Curie constant $C(T)=T\chi(T)$
are shown in Fig.~\ref{fig:CSSE}. At the highest temperatures
the effective dopant spins behave as free spins while at low temperature
we observe a new Curie-like behavior with a reduced Curie constant
$\sim 1/12$. Although here the 2D system orders
at $T=0$ (as proven above)
this behavior agrees with a qualitative argument based on the formation of large spin clusters due to the presence of AF as well as ferromagnetic (F) couplings, applied 
by Sigrist and Furusaki~\cite{note2} in the 1D case.
\subsection{Real Space Renormalization Group}
In order to explore this large spin formation, we follow the pioneering 
work of Ma, Dasupta and Hu~\cite{MDH79} and also Bhatt and Lee~\cite{BL82},
and extend the Real Space Renormalization Group (RSRG) scheme to hamiltonian (\ref{EffHam})
with F and AF long-distance couplings. It turns out that such a RSRG procedure 
has allowed the identification of the so-called {\it{Infinite Randomness Fixed point}} (IRFP) by D.~Fisher for the disordered 
Heisenberg spin-$\frac{1}{2}$ chain~\cite{Fisher94} as well as for the Ising chain in transverse field at criticality~\cite{Fisher95}.
Nevertheless, such an IRFP (as well as its associated {\it{random Singlet Phase}}) is unstable against the introduction of randomness in the couplings' signs, as shown first by Westerberg {\it et~al.}~\cite{Sigrist1} for the Heisenberg spin-$\frac{1}{2}$ 
chain with random couplings in magnitude {\it and} sign. 
The new fixed point identified in the 1D case, using an extended RSRG scheme~\cite{Sigrist2} and checked numerically using QMC simulations~\cite{MonteCarloFAF}, is especially characterized by a large spin formation~\cite{noteLSP}. 
%
\begin{figure}[!ht]
\bc
\epsfig{file=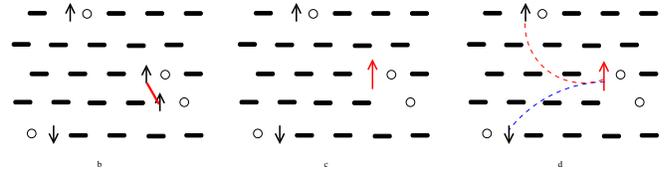,width=\columnwidth,clip}
\caption{Schematic picture of one RG step performed in the doped spin-Peierls system.
Thick bonds stand for dimers and non-magnetic impurities
(released spins $\frac{1}{2}$) are represented by open circles
(black arrows). The initial RSRG step is illustrated starting from
a typical configuration with 4 impurities:
(a) The strongest coupled pair is identified (red line).
(b) This pair, e.g. ferromagnetic here, is
replaced by a spin $S=1$ (red arrow). (c) The
couplings with all other spins (dashed lines) are renormalized.}
\label{fig:PictRSRG} \ec
\end{figure}
%

In the 2D problem adressed here, long range interactions with random sign and magnitude can also be studied by an extended version of the RSRG technique.
Let
us first define the effective interaction
as $J_{i,j}$ where $i$ and $j$ label the {\it randomly distributed
spins} and run from $1$ to $N_s$.  One single RG step is described
as follows (see Fig.~\ref{fig:PictRSRG}): 1) Identify the most strongly coupled pair of spins
($S_1,S_2$) i.e. with the largest energy gap $\Delta_{1,2}$, $
\Delta_{1,2}=J_{1,2}(1+\left|S_1 -S_2 \right|)
~{\rm{if}}~J_{1,2}>0~{\rm{(AF)}}$ and $\Delta_{1,2}=-J_{1,2}(S_1
+S_2) ~{\rm{if}}~J_{1,2}<0~{\rm{(F)}}$. Note that $\Delta_{1,2}$
defines the energy scale of the transformation which will, in fact, play the role of the temperature.
2) Replace it by an
effective spin $S^{'}=\left|S_1 -S_2 \right|$ if the coupling is
AF or $S^{'}=S_1 +S_2$ in the F case. 3) Renormalize all the
magnetic couplings with the following rules : (i) If $S^{'}\neq
0$, as given by a first order perturbation theory, the new
couplings between $S^{'}$ and all the other spins ($S_3,
S_4,...,S_{N_s}$) are set to
\be
{\tilde{J}}_{(S^{'},S_{i})}=J_{1,i}~c(S_1,S_2,S^{'})+J_{2,i}~c(S_2,S_1,S^{'}),
\ee
with
\be
c(S_1,S_2,S^{'})=\frac{S^{'}(S^{'}+1)+S_1(S_1+1)-S_2(S_2+1)}{2S^{'}(S^{'}+1)}.
\ee
(ii) If $S^{'}=0$, the pair ($S_1,S_2$) is frozen.
Using a cluster approximation~\cite{BL82}
that involves only the extra pair ($S_3,S_4$)
 the most strongly coupled to
$S_1$ and $S_2$ and a second order perturbation,
the coupling $J_{3,4}$ is renormalized as
\be
{\tilde{J}}_{3,4}=J_{3,4}+\frac{2S_1(S_1+1)}{3J_{1,2}}(J_{1,3}-J_{2,3})(J_{2,4}-J_{1,4}).
\ee
The same procedure is then reiterated. We also check that the RSRG preserves the non-frustrated character of the problem.
\subsection{Large spin formation}
Due to the presence of both F and AF couplings, clusters with
large effective spins are created during the procedure similarly
to what occurs in the 1D random F-AF spin-$\frac{1}{2}$
chain~\cite{Sigrist1}. At each RG step, the energy scale
$\Delta_0$ decreases and both the number of inactive spins frozen
into singlets and the number of clusters build from a large number
$n$ correlated spins-$\frac{1}{2}$, increase. The aforementioned random
walk argument predicts that, the average number $\langle n
\rangle$ of spins-$\frac{1}{2}$ inside clusters and their average spin
magnitude $\langle S^{\rm{eff}}\rangle $ should be related by
$\langle S^{\rm{eff}}\rangle \sim \langle n \rangle ^{1/2}$ at low
enough temperatures. 
%
\begin{figure}[!ht]
\bc
\psfrag{S}{$\langle S^{\rm{eff}} \rangle$}
\psfrag{D}{$\langle \Delta_0 \rangle$}
\psfrag{n}{\tiny{$\langle n \rangle$}}
\psfrag{d}{\tiny{$\langle \Delta_0 \rangle$}}
\epsfig{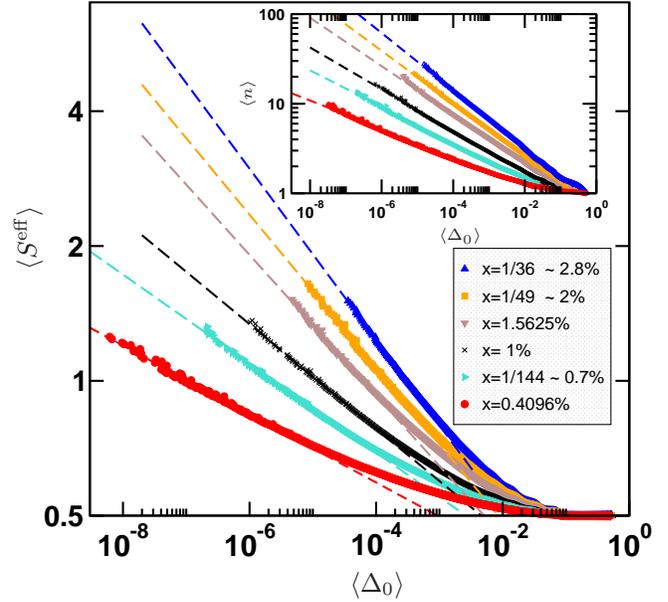}
\caption{Average effective spin $\langle S^{\rm{eff}}\rangle$ of the
clusters of active spins vs the energy scale $\langle \Delta_0
\rangle$ for six different concentrations $x$ indicated on the
plot. Numerical RSRG data obtained for $N_s =1024$ spins over more
than $10^4$ samples. Inset: for the same samples, average number
$\langle n \rangle$ of initial spins-$\frac{1}{2}$ per cluster vs
$\langle\Delta_0 \rangle$. Dashed lines are power-law fits (see
text). (Figure reprinted from Ref.\cite{LPS05}).} \label{fig:SpinEffRSRG} \ec
\end{figure}
Therefore we expect the effective spin of
these clusters to grow monotonously as the energy scales down. We
have analyzed this process using the RSRG scheme to compute both
$\langle S^{\rm{eff}}\rangle $ and $\langle n \rangle$ as a
function of $\langle \Delta_0\rangle$. This is shown in
Fig.~\ref{fig:SpinEffRSRG} which clearly demonstrates the
formation of large moments. Moreover, power-law divergences like
\be
\langle S^{\rm{eff}}\rangle \sim \langle\Delta_0 \rangle
^{-\alpha (x)}~~~{\rm{and}}~~~\langle n \rangle \sim \langle\Delta_0
\rangle ^{-\kappa (x)}
\ee
are observed with $\kappa \simeq 2
\alpha$. 
Interestingly enough, we find that $\alpha$
depends on $x$, like $\sim \sqrt{x}$, in contrast to the random F-AF spin chain for which
$\alpha=0.22\pm 0.01$~\cite{Sigrist2,MonteCarloFAF,noteRG}. 

\begin{figure}[!ht]
\bc
\psfrag{C}{${\overline{c}}({\overline{\Delta_0}})$}
\psfrag{D}{${\overline{\Delta_0}}$}
\epsfig{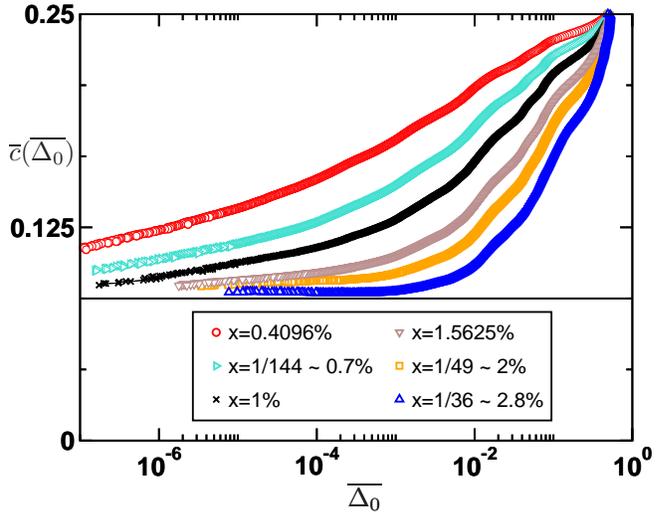}
\caption{Curie constant per spin $\langle C \rangle$ plotted vs the energy scale.
Numerical RSRG results shown for
$N_s =1024$ spins and 6 different concentrations $x$ as indicated
on the plot, vs the RG energy scale $\langle\Delta_0 \rangle$. Error
bars are typically smaller than symbol sizes, the number of
samples always exceeding $10^4$.  The full line correspond to the
saturation value of $1/12$. (Figure reprinted from Ref.\cite{LPS05}).}
\label{fig:CurieRSRG}
\ec
\end{figure}

\begin{figure}[!ht]
\bc
\psfrag{C}{$\langle C \rangle-\frac{1}{12}$}
\psfrag{A}{\tiny{(a) $x\simeq 0.41\%$}}
\psfrag{B}{\tiny{(b) $x\simeq 0.7\%$}}
\psfrag{D}{\tiny{(c) $x=1\%$}}
\psfrag{E}{\tiny{(d) $x\simeq 1.56\%$}}
\psfrag{F}{\tiny{(e) $x\simeq 2\%$}}
\psfrag{G}{\tiny{(f) $x\simeq 2.8\%$}}
\psfrag{T}{$T,~\langle \Delta_0 \rangle$}
\epsfig{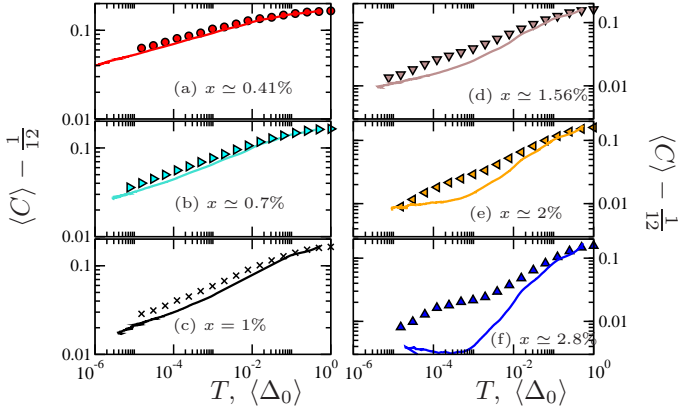}
\caption{Direct comparisons
between RSRG (full lines) and SSE simulations (symbols) of $\langle C
\rangle-1/12$ are shown for the 6 different concentrations indicated by 
(a), (b),..., (f) vs the RG energy scale $\langle\Delta_0 \rangle$ or 
the SSE temperature $T$. (Figure reprinted from Ref.\cite{LPS05}).}
\label{Compare} \ec
\end{figure}

The Curie constant can also be evaluated using the decimation procedure.
The RSRG computation of $C$ is
performed, at each RG step, using the formula
$C=\frac{1}{3N_s}\sum_{\sigma}N_{\sigma}\sigma (\sigma +1)$ where
$N_{\sigma}$ is the number of active effective spins of size
$\sigma$, the data being then averaged over disorder. We have also
checked that finite size effects are negligible when $N_s \ge 256$
and we have chosen $N_s =1024$ in most computations.
Fig.~\ref{fig:CurieRSRG} shows results for the Curie constant per spin
obtained with the decimation procedure. 
We can confront the RSRG results  to QMC data obtained on the same
model and for the same parameters.
%
We observe a
qualitative agreement between Fig.\ref{fig:CSSE} and
Fig.\ref{fig:CurieRSRG} where, at high temperatures, the spins behave
as paramagnetic free magnetic moments (giving a Curie constant of
$\frac{1}{4}$ per spin) and where saturation to $\frac{1}{12}$ is
observed at low $T$, the spins being correlated inside large
clusters. At small concentration, the agreement becomes even
quantitative, as seen in the Fig.\ref{Compare} where a direct comparison between RSRG 
and SSE is shown for the six lowests values of $x$.
%
%
\subsection{Scaling regime}
\begin{figure}[!ht]
\bc
\psfrag{C}{$T\times \chi_{\rm{stag}}$}
\psfrag{T}{$T$}
\epsfig{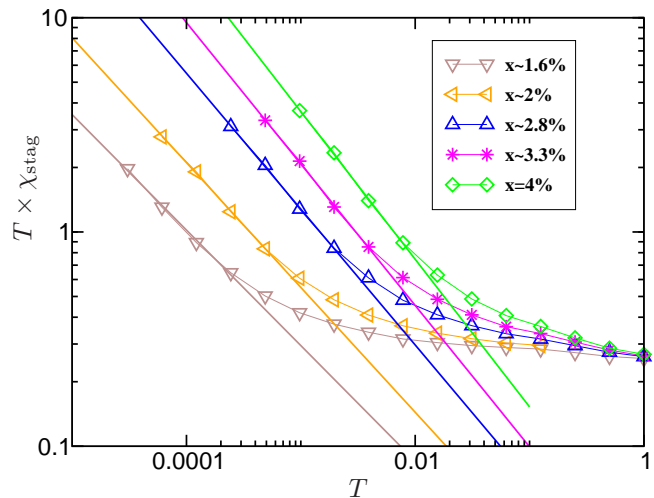}
\caption{$T\langle \chi_{\rm stag}(T)\rangle$ plotted vs
$T$ for five different concentrations. Full
lines are fits corresponding to power-law behaviors $\sim T^{-2\gamma'}$. All data are
computed by QMC using $N_s=256$ random spins and averaged over
disorder. (Figure reprinted from Ref.\cite{LPS05}).} \label{fig:ChiStag} \ec
\end{figure}

\begin{figure}[!ht]
\bc
\psfrag{A}{{\Large{$x$}}}
\psfrag{cr}{{\bf{Critical exponents}}}
\epsfig{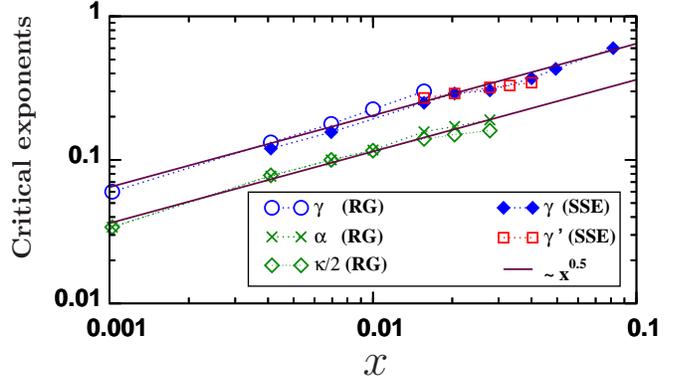}
\caption{Exponents
$\alpha(x)$, $\gamma(x)$ and $\gamma'(x)$ extracted from various SSE and RSRG data 
as indicated on the plot (see text for details). Straight lines are $\sim \sqrt{x}$. (Figure reprinted from Ref.\cite{LPS05}).}
\label{fig:alphax}
\ec
\end{figure}

We now turn to the analysis of the scaling regime. At very low
temperatures, from the analogy with 1D systems we expect the following quantum corrections 
\be
\langle C(T)\rangle -1/12 \sim T^{\gamma},
\ee
with $\gamma=\alpha$ in 1D~\cite{MonteCarloFAF}.
Similarly, one expects for the staggered
susceptibility (per spin), 
\be
\langle \chi_{\rm{stag}}(T)\rangle \sim T^{-1-2\gamma'}.
\ee
In 1D, $\gamma'$ is expected to be equal to $\alpha$~\cite{Nagaosa96},
but in the present case, by direct fits of the low T (see Fig.~\ref{fig:ChiStag}), we found 
$$\gamma'\simeq\gamma\simeq 2\alpha.$$
%
%
Thoses exponents are plotted in
Fig.~\ref{fig:alphax}. An overall very good agreement is seen
between the different methods, in particular between the estimates of $\gamma$
obtained from the analysis of the Curie constant computed by QMC
and RSRG. We stress again that the exponent $\alpha$ deduced from the analysis of
the change of cluster sizes and spins is roughly a factor of
2 smaller than $\gamma$. Interestingly enough,
$\gamma \simeq 2\alpha \propto \sqrt{x}$ in all cases.

It is remarkable
that such properties are observed above the
$T=0$ ordered magnetic groundstate. Moreover, it is the first example of a two-
dimensionnal random magnet exhibiting a large spin phase with a disorder (the 
concentration $x$) dependence of the critical exponents. Whereas doped CuGeO$_3$
 might be a good candidate for the experimental observation of a critical regime, 
the so large three-dimensionnal ordering temperature prevents such an observation. 
More strongly diluted samples are necessary to reach the scaling regime.

\section{Conclusion}

Starting from a microscopic model for impurity-doped
dimerized and frustrated coupled spin-chains (similar to let's say, Zn-doped
copper germanate), we have derived a simple and tractable effective 
model describing S=1/2 effective spins interacting through 
long range (spatially anisotropic) interactions. An important property of this
long-range exchange is its oscillating nature (with both AF and F couplings) 
in such a way it is essentially non-frustrating, hence leading to an AF 
instability at low temperature 
(in the presence of a small 3D coupling). This is a remarkable fact 
since magnetic ordering in the pure system is prevented by frutration.
Using state-of-the-art SSE QMC computations thermodynamic properties of the 
effective model are studied in details down to very low temperatures.
Critical behaviors with impurity concentration-dependent exponents 
are found and interpreted using a new Real Space RG 
procedure clearly establishing the important role of large clusters of 
correlated spins carrying a finite magnetization.

\section*{Acknowledgment}
We are deeply grateful to A.W.~Sandvik and M.~Sigrist for many useful discussions and 
some collaborations. 
We also thank IDRIS (Orsay, France) for allocation of CPU time on their supercomputers.
N.L. thanks I.~Affleck for hospitality at UBC. 

\appendix


\begin{thebibliography}{99}
\bibitem{NoteRG} Note however that the
universality class identified in \cite{Sigrist1} as a RG fixed
point occurs only for initial gap distributions less singular than
$P_c (\Delta)\sim \Delta ^{-y_c}$, with $0.65\lesssim y_c \lesssim
0.75$. For more singular distributions, as it is the case here,
critical exponents are not universal anymore, at least in
1D~\cite{Sigrist2}.
\end{thebibliography}
\end{document}